\documentclass[twocolumn]{aastex62}

\newcommand\aastex{AAS\TeX}

\newcommand{\ie}{{\it i.e.}}
\newcommand{\eg}{{\it e.g.}}
\newcommand{\goes}{{\it GOES}}
\newcommand{\sdo}{{\it SDO}}
\newcommand{\rhessi}{{\it RHESSI}}
\newcommand{\fermi}{{\it FERMI}}

\newcommand{\aia}{{\it AIA}}

\accepted{by ApJ Letters on November 25, 2019}

\shorttitle{\aastex\ sample article}
\shortauthors{Hernandez-Perez et al.}

\begin{document}

\title{A HOT CUSP-SHAPED CONFINED SOLAR FLARE}

\correspondingauthor{Aaron Hernandez-Perez}
\email{aaron.hernandez-perez@uni-graz.at}

\correspondingauthor{Yang Su}
\email{yang.su@pmo.ac.cn}

\author{Aaron Hernandez-Perez}
\affil{University of Graz, Institute of Physics/IGAM, A-8010 Graz, Austria}

\author{Yang Su}
\affil{Key Laboratory of Dark Matter and Space Astronomy, Purple Mountain Observatory\\
Chinese Academy of Sciences, 8 Yuanhua Road, Nanjing 210034, China}

\author{Julia Thalmann}
\affil{University of Graz, Institute of Physics/IGAM, A-8010 Graz, Austria}

\author{Astrid M. Veronig}
\affil{University of Graz, Institute of Physics/IGAM, A-8010 Graz, Austria}

\author{Ewan C. Dickson}
\affil{University of Graz, Institute of Physics/IGAM, A-8010 Graz, Austria}

\author{Karin Dissauer}
\affil{University of Graz, Institute of Physics/IGAM, A-8010 Graz, Austria}

\author{Bhuwan Joshi}
\affil{Udaipur Solar Observatory, Physical Research Laboratory, Udaipur 313 001, India}

\author{Ramesh Chandra}
\affil{Department of Physics, DSB Campus, Kumaun University, Nainital 263 002, India}

\begin{abstract}

We analyze a confined flare that developed a hot cusp-like structure high in the corona (H$\sim$66~Mm). A growing cusp-shaped flare arcade is a typical feature in the standard model of eruptive flares, caused by magnetic reconnection at progressively larger coronal heights. In contrast, we observe a static hot cusp during a confined flare. Despite an initial vertical temperature distribution similar to that in eruptive flares, we observe a distinctly different evolution during the late (decay) phase, in the form of prolonged hot emission. The distinct cusp shape, rooted at locations of non-thermal precursor activity, was likely caused by a magnetic field arcade that kinked near the top. Our observations indicate that the prolonged heating was a result of slow local reconnection and an increased thermal pressure near the kinked apexes due to continuous plasma upflows.\\

\end{abstract}


\section{Introduction}
     \label{S-Introduction} 


\begin{figure*}[ht]
\centerline{
\centering\includegraphics[width=\textwidth]{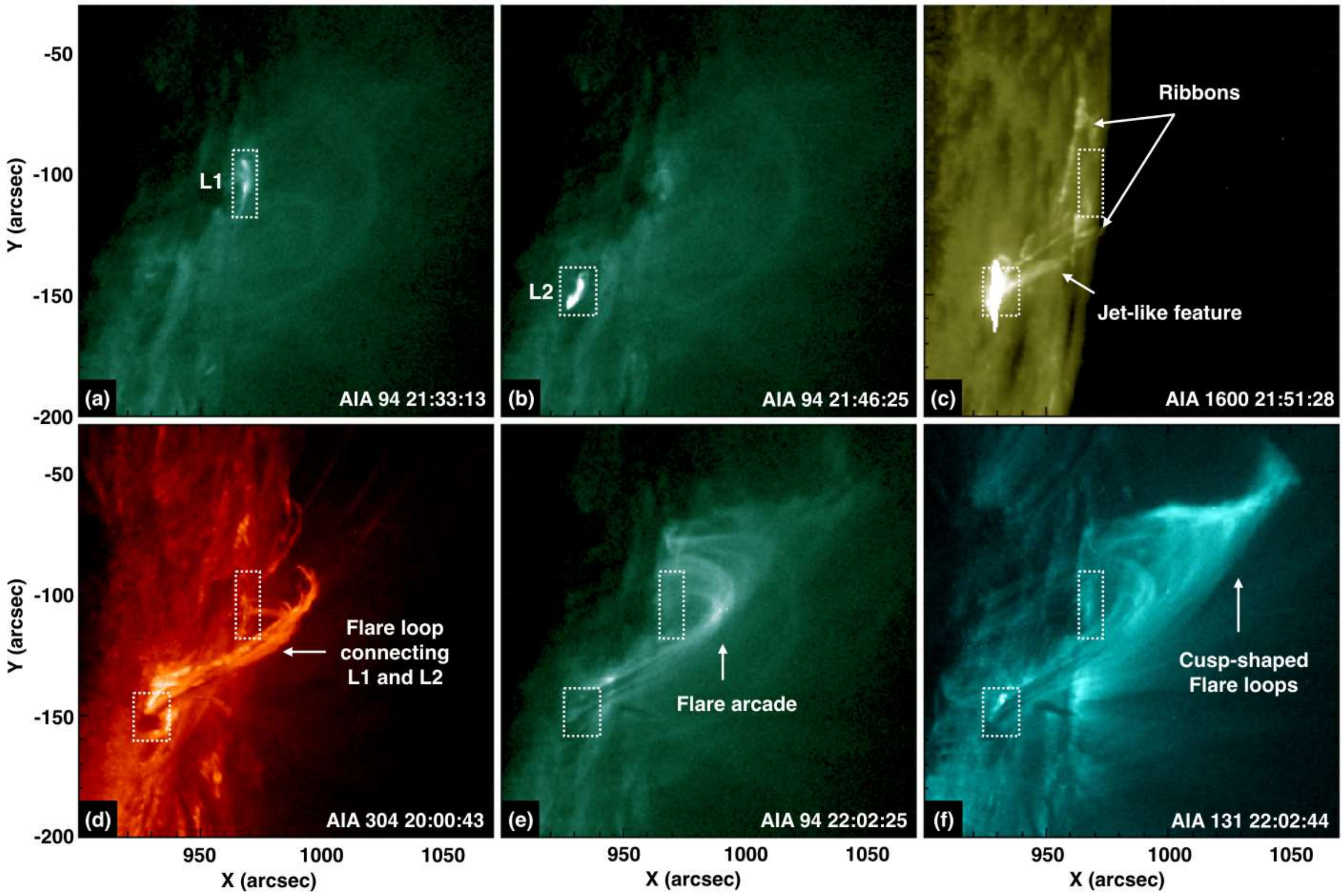}
}
\caption{(E)UV sequence showing the main aspects of SOL2014-01-13T21:51M1.3 in different \aia\ filters. (a,b) EUV precursor brightenings at L1 and L2. (c) Formation of ribbons and jet-like feature during the impulsive phase. (d--f) EUV sequence of the decay phase, showing the connectivity between the precursor sites and the cusp-shaped flare loops. The animation shows the evolution of the flare in co--temporal AIA~1600, 304, 94 and 131~\AA\ maps.\\
(An animation of this figure is available.)
  }
\label{EUV_summary}
\end{figure*}


\begin{figure*}[ht]
\centerline{
\centering\includegraphics[width=\textwidth]{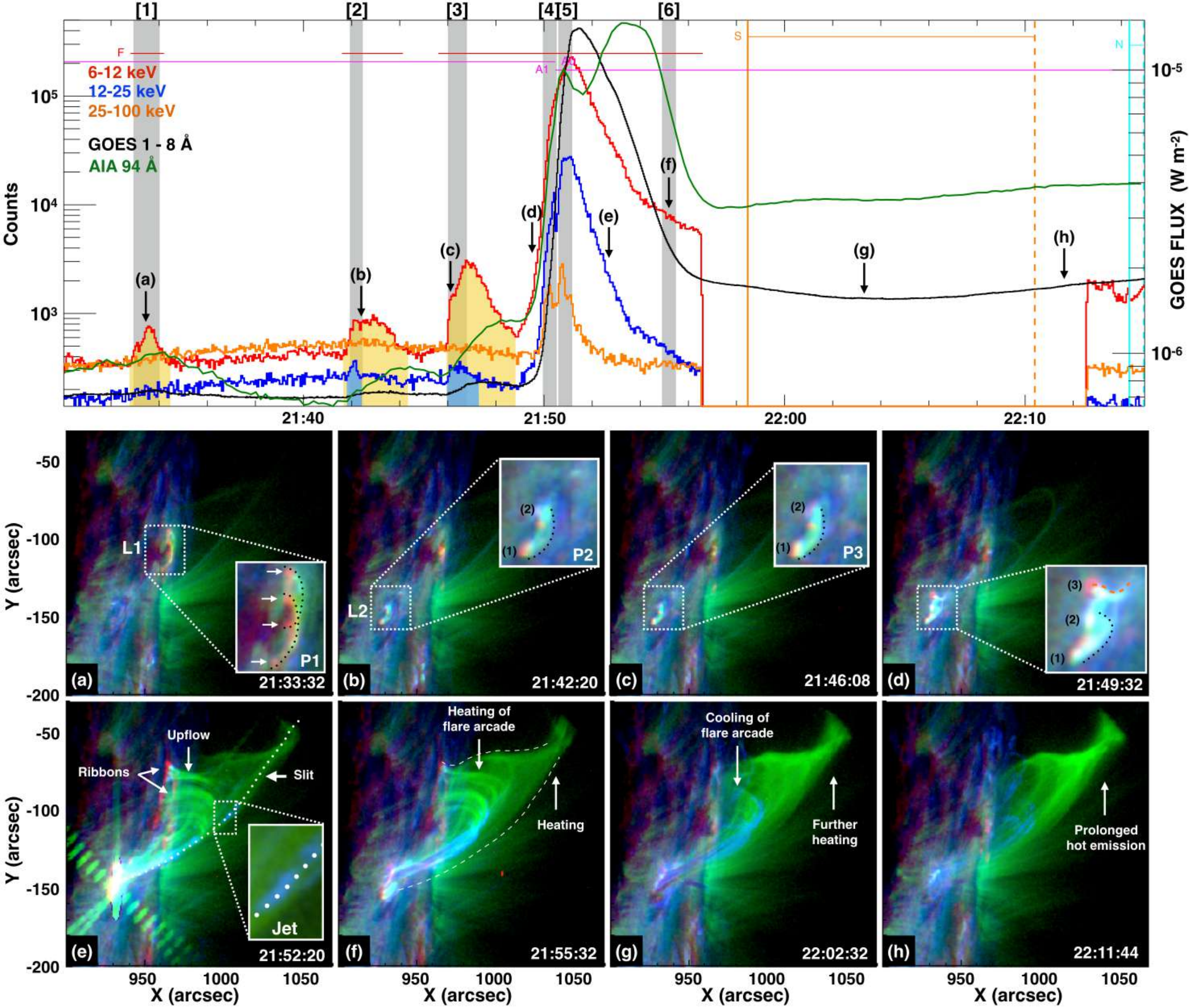}
}
\caption{
Top: \rhessi\ X-ray counts, for the 6--12~(red), 12--25~(blue) and 25--100~keV~(orange) energy bands, \goes\ SXR flux in the 1--8~\AA\ wavelength band (black solid line), and normalized \aia~94~\AA\ lightcurve for the FOV in the bottom pannels (green solid line). The yellow and blue areas highlight the enhanced X-ray precursor emission in the 6--12 and 12--25~keV energy bands respectively. The grey areas [1--6] indicate the integration times for which the \rhessi\ spectra in Fig.~\ref{spectra} were derived. Bottom: composites of \aia~1600~(red)~+~304~(blue)~+~131~\AA~(green) images showing the (E)UV flare emission. The times are indicated on the top panel by letters ((a)--(h)) corresponding to each of the composites.\\
(An animation of this figure is available.)
  }
\label{Lightcurve}
\end{figure*}

\begin{figure*}[ht]
\centerline{
\centering\includegraphics[width=\textwidth]{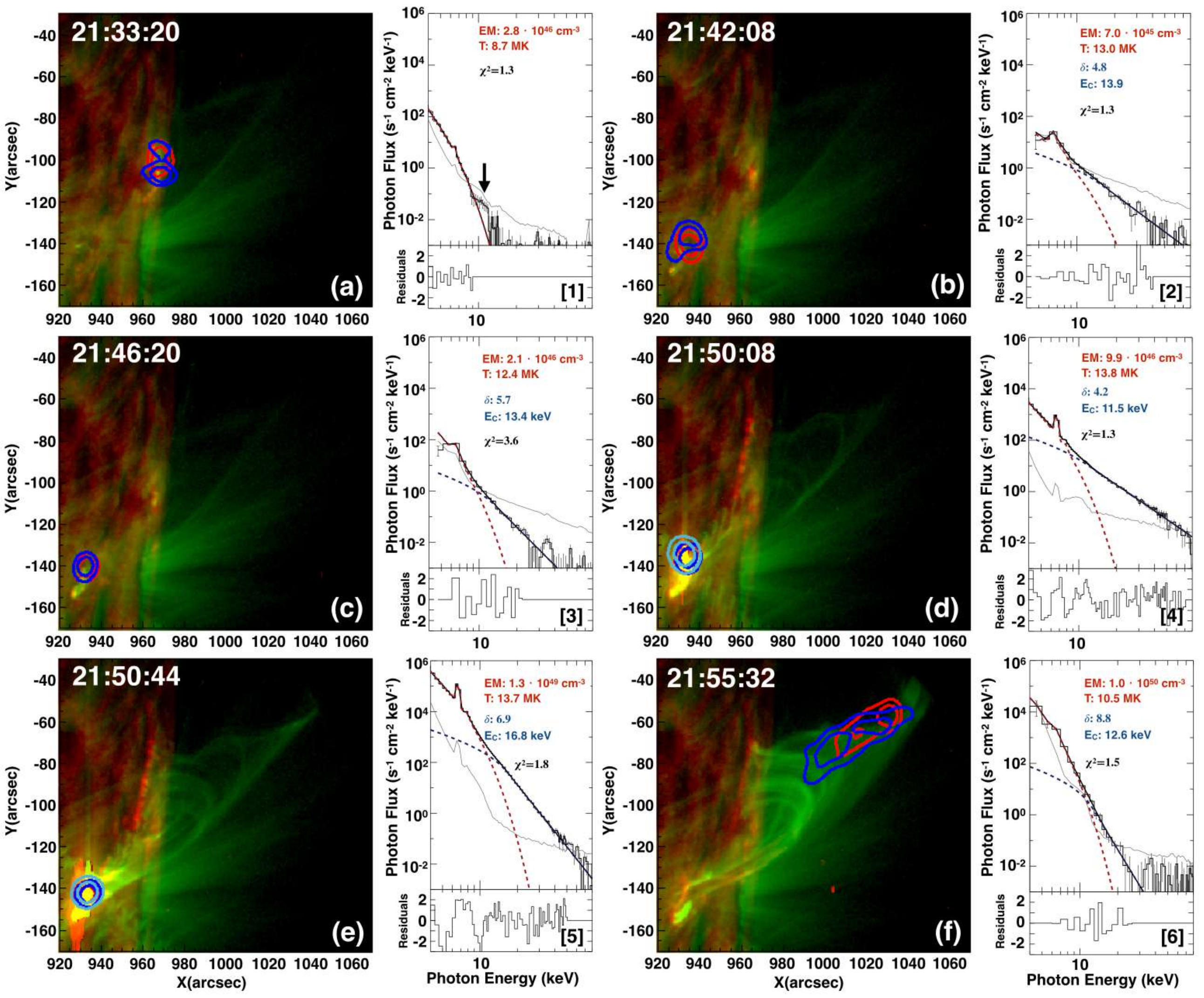}
}
\caption{(a--f) Composites of \aia~1600~(red)~+~131~\AA~(green) images showing the (E)UV flare emission. The \rhessi\ sources for 3--8 (red), 8--20 (dark blue), and 20--100~keV (light blue) with contours at 60\% and 80\% of the maximum emission are overplotted. [1--6] corresponding X-ray spectra for the integration times represented by the grey areas in the top panel of Fig.~\ref{Lightcurve}. All spectra correspond to \rhessi\ with exception of [2] and [3], which show \fermi\ spectra because they provided better statistics. We note that the corresponding \rhessi\ spectra at those times showed very similar results. The X-ray spectra of background-subtracted data (black solid lines) are plotted together with the fitting results for the isothermal component (red dashed lines), and the non-thermal component (blue dashed lines). The background is represented by the grey solid line. The electron temperature, $T$, emission measure, $EM$, electron distribution index, $\delta$, and cutoff energy, $E_{C}$, as well as the chi-squared of the fitting, ${\chi}^2$, are listed.
  }
\label{spectra}
\end{figure*}

Solar flares are explosive events in which magnetic energy is rapidly converted into plasma heating and particle acceleration \cite[\eg][]{2017LRSP...14....2B}. The most accepted model for eruptive flares is the so-called CSHKP model \citep[\eg][]{1964NASSP..50..451C,1966Natur.211..697S,1974SoPh...34..323H,1976SoPh...50...85K}. In the simplistic view of this 2D framework, an erupting flux rope stretches the magnetic field underneath, forming a current sheet, toward which the embedding field is drawn and forced to reconnect \citep[][]{2000JGR...10523153F}. The ability to explain many features observed in eruptive flares is where the success of this model resides. For instance, hot cusp-shaped coronal structures often observed in flares are regarded as observational evidence of the reconnection process \citep{1992PASJ...44L..63T}. These structures exhibit an increasing temperature with altitude, caused by the newly reconnected fields, which are drawn back from the reconnection site \citep{1996ApJ...464.1055T}. As reconnection occurs progressively higher up in the corona, the flare arcade grows, accompanied by a separation in the quasi-parallel ribbons, establishing the connectivity between the newly reconnected loops and the chromosphere \citep[\eg,][]{2006A&A...446..675V}.

Despite the ability of the CSHKP model to describe several features in eruptive flares, a significant fraction of the flares are confined (\ie\ they are not associated with the eruption of a flux rope). Confined flares cannot be accommodated in the standard flare model, and other theories are needed to understand their physics. Several models have been proposed including, \eg, quadrupolar current-loop interaction \citep[\eg][]{1997ApJ...486..521M}, emerging flux \citep{1977ApJ...216..123H}, or the fan-spine topology \citep[\eg][]{2009ApJ...700..559M}. These models can explain features of confined flares that deviate from those observed of eruptive flares, \eg, the interaction between current-carrying loops \citep[\eg][]{2000ApJ...540.1143Y,2017ApJ...845..122G}, the interaction of emerging flux and coronal fields \citep[\eg][]{2015SoPh..290.2923V}, and the formation of circular or quasi-circular ribbons \citep[\eg][]{2012A&A...547A..52R,2015ApJ...812...50J,2017ApJ...847..124H,2018ApJ...859..122L,2019ApJ...871..105Z,2019ApJ...871....4H,2019ApJ...878...78C,2019ApJ...883..124Z}.

However, although rarely reported in literature, some confined flares occasionally exhibit features similar to those observed in eruptive flares. For instance, observational evidence of cusp-shaped loops have been reported on two occasions. \cite{2014ApJ...790....8L} reported a confined flare exhibiting a diffusive cusp-shaped structure, which was interpreted as a result of the sudden changes in the magnetic field connectivity that reconnected across a quasi-separatrix layer (QSL). However, unlike in eruptive flares, the temperature in the cusp was lower than the underlying flare arcade. \cite{2015SoPh..290.2211G} presented observations of a confined event (event No.6 in their sample) that exhibited a double candle-flame configuration, \ie\ two cusp-shaped structures located side by side, that shared the cusp-shaped edges. They observed a temperature distribution similar to those typically observed in eruptive events, and interpreted this phenomenon as heating from the slow-mode shocks from the reconnection site.

We report an atypical confined flare (SOL2014-01-13T21:51M1.3) that exhibited an apparent cusp with a temperature distribution otherwise usually only observed for eruptive flares. The decay phase was characterized by unusual prolonged hot emission, originating from the cusp's apex. This study presents a new scenario in the initiation of confined cusp-shaped flares and aims to elucidate the mechanism responsible for the extended heating.

\section{Data \& Methods}
     \label{S-Data}

We used (extreme) ultra-violet ((E)UV) data from the Atmospheric Imaging Assembly \citep[\aia;][]{2012SoPh..275...17L} on board the {\it Solar Dynamics Observatory} \citep[\sdo;][]{2012SoPh..275....3P} to study the evolution of the flare plasma. To study the nonthermal signatures of accelerated electrons as well as the hot thermal flare emission, we used hard X-ray data from the {\it Ramaty High Energy Spectroscopic Imager} \citep[\rhessi;][]{2002SoPh..210....3L} and the \fermi\ satellite \citep{2009ApJ...702..791M}. \rhessi\ CLEAN images \citep{2002SoPh..210...61H} were constructed using the front segments of detectors 1, 3, 4, 5, 6, 8, and 9. X-ray spectra were fitted with an isothermal model and, when appropriate, a thick--target non--thermal emission model \citep{1971SoPh...18..489B,2003ApJ...595L..97H}. For the spectral analysis, we selected the detectors that provided the best fitting results, namely 4 and 6 for \rhessi. For \fermi{\it /GBM}, detector 5 was used, as it was pointing to the Sun during the flare.

The thermal evolution of the flare plasma was studied by means of a Differential Emission Measure (DEM) analysis on \aia\ EUV filtergrams. We used the Sparse inversion method developed by \cite{2015ApJ...807..143C}, with the new settings proposed by \cite{2018ApJ...856L..17S}.

\section{Results}
     \label{S-Results}

\subsection{Event Overview}
     \label{S-Event Overview}

The confined M1.3--class flare on 2014 January 13 occurred in NOAA AR 11944 near the western solar limb (S10W81). Fig.~\ref{EUV_summary} and animation 1 show the main (E)UV aspects of SOL2014-01-13T21:51M1.3. The early phase was characterized by localized (E)UV enhancements (\ie\ precursors) at two locations very low in the corona (marked as L1 and L2 in Fig.~\ref{EUV_summary}(a,b)), separated by a projected distance of $\sim$37~Mm. The triggering of the flare produced a jet--like feature and ribbons encompassing L1 (Fig.~\ref{EUV_summary}(c)). The direct connectivity between the precursor sites was observed by hot flare loops connecting L1 and L2 during the flare decay phase (Fig.~\ref{EUV_summary}(d)). The EUV emission during the decay phase revealed the flare arcade, and the hot cusp-shaped flare loops (Fig.~\ref{EUV_summary}(e,f)), extending up to $\sim$66~Mm above the solar surface (distance measured from L1 to the highest point). Prolonged hot emission of the cusp-shaped flare loops followed (see animation 1).

In order to understand the formation of this unusual coronal structure and its subsequent prolonged hot emission, we need to understand the chain of events that led to the flare and its underlying magnetic configuration. The following sections present a detailed description and analysis of the flare characteristics, chronologically addressing: 1. the flare precursors, 2. the causal relationship between precursors and flare features, and 3. the characteristics of the cusp-shaped flare loops and the prolonged hot emission.

\subsection{SXR precursors}
     \label{S-Precursors}

Flare precursors are enhanced pre-flare soft X-ray (SXR) emission indicative of small-scale energy release that may play a role in the triggering of the main flare \citep[\eg][]{1984AdSpR...4...95V,2002SoPh..208..297V,2016ApJ...832..130J,2019ApJ...874..122H}. Three precursor episodes (\goes--class B7.4, B7.0 and B7.9) were registered by \goes\ and \rhessi\ during the early phase of the M1.3 flare under study (see enhanced 6--12 and 12--25~keV precursor emission highlighted by the yellow and blue areas in the top panel of Fig.~\ref{Lightcurve}). Each SXR precursor was co--temporal with a localized (E)UV enhancement, occurring at two well separated locations very compact and low in the corona, \ie\ L1 and L2 (see Fig.~\ref{Lightcurve}(a--c) and animation), with a projected distance of $\sim$37~Mm. (E)UV observations of the precursors (see P1, P2 and P3 in the insets of Fig.~\ref{Lightcurve}(a--c)) show typical signatures indicative of energy release, \ie\ bright loops (indicated by the black dotted lines) connecting chromospheric brightenings (marked by white arrows for P1 and labeled (1,2) for P2 and P3). 

The corresponding X-ray imaging and spectroscopy show compact sources co-spatial with the precursor sites and weak, yet significant, episodes of enhanced emission (see Fig.~\ref{spectra}(a--c)). Due to low statistics in the X-ray spectra associated with Fig.~\ref{spectra}(a), a non-thermal component was not included in the fitting. However, at energies $>$10~keV, a clear enhancement in the photon flux above the thermal fit, was registered (marked by a black arrow in Fig.~\ref{spectra}(a)), indicative of a weak non-thermal tail. The spectrum associated with P2 and P3 (see Fig.~\ref{spectra}(b,c)) showed significant non--thermal emission at these times, co--temporal with an increase in the \aia~94~\AA\ emission during the rest of the early phase (see the \aia~94~\AA\ lightcurve in Fig.~\ref{Lightcurve} at 21:42--21:49~UT). The DEM analysis revealed pre-heating episodes at the precursor locations (see Fig.~\ref{DEM}(a) and animation), accounting for thermal plasma at a temperature of $\sim$8~MK. The triggering of the flare immediately followed P3. 

In order to infer information on the triggering of the event and its magnetic topology, we initially studied the magnetic field of AR 11944 during the flare under study as well as a few days before. We note that due to the rapidly varying photospheric magnetic field of the AR prior to the occurrence of SOL2014-01-13T21:51M1.3, its closeness to the limb and the small scale nature of the event, the magnetograms did not provide us with useful information for the magnetic configuration on the date of the flare occurrence. 

\subsection{Causal relationship of precursors and flare features}
     \label{S-Causal relationship of precursors and flare features}

In addition to the bright loop and associated chromospheric brightenings observed for P2 and P3, the impulsive flare phase starts with the appearance of an additional loop (marked by the orange dashed line in the inset of Fig.~\ref{Lightcurve}(d)) rooted at a chromospheric brighening (labeled (3)). This was immediately followed by the generation of flare ribbons neighboring L1 and the enhanced emission at L2 as a consequence of non-thermal bremsstrahlung (Fig.~\ref{Lightcurve}(e)). No X-ray sources were found at the locations where the ribbons formed encompassing L1 (Fig.~\ref{spectra}(d,e)), possibly due to \rhessi's limited dynamic range. However, strong non-thermal emission during the impulsive flare phase was registered at L2, in the form of a compact HXR source and a flat non-thermal ($\delta\sim4.2$) component in the corresponding spectrum (Fig.~\ref{spectra}(d)). This shows that L2 was the main compact flare region. 

\begin{figure*}[ht]
\centerline{
\centering\includegraphics[width=0.75\textwidth]{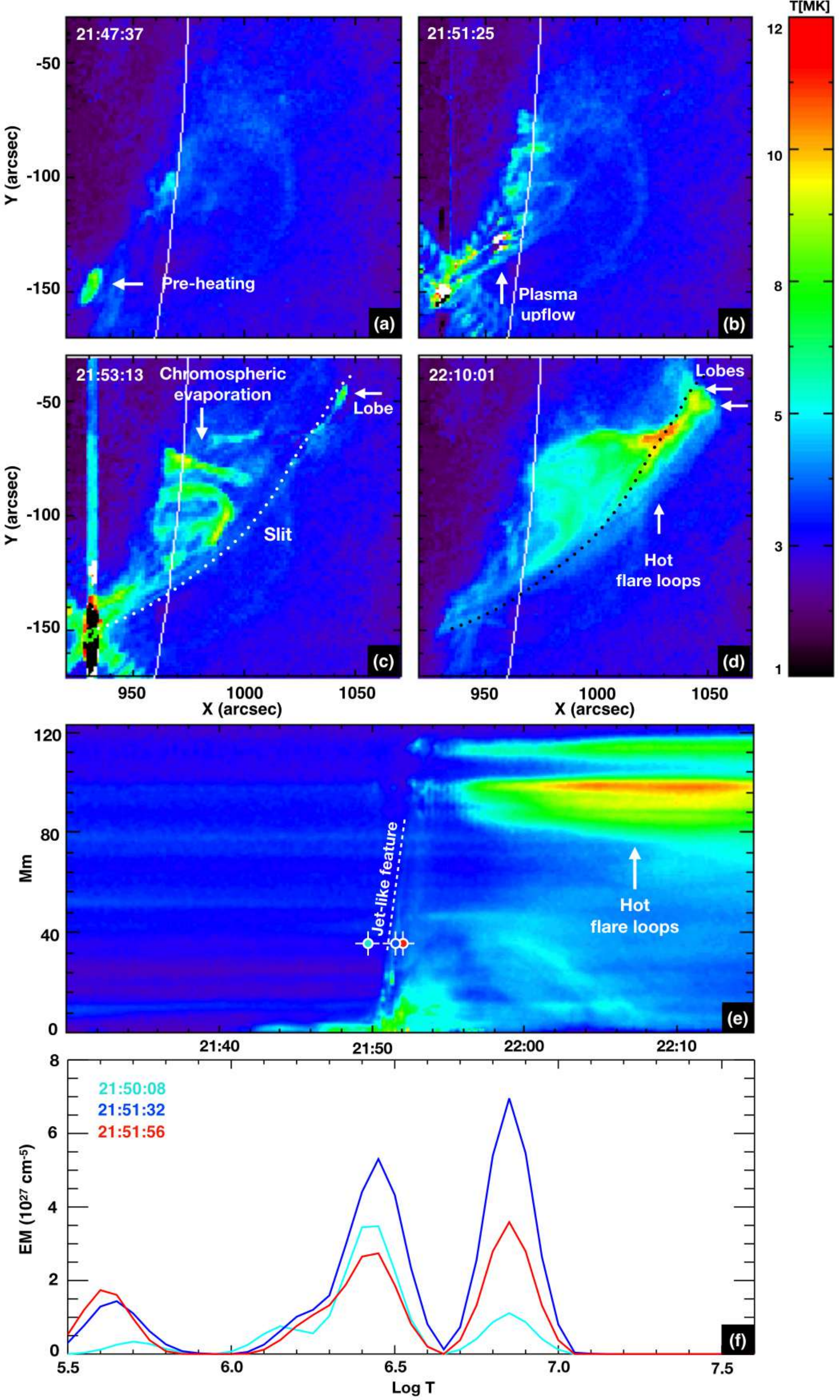}
}
\caption{
(a--d) EM weighted temperature maps during the M1.3 flare. The dotted lines represent an artificial slit along the trajectory of the jet-like feature, along which we extracted the DEM evolution.; (e) Time-distance plot for EM-weighted temperatures along the slit.; (f) EM distribution for the position and times marked by the colored circles in panel (e).\\ 
(An animation of this figure is available.)
  }
\label{DEM}
\end{figure*}

During this phase, a jet--like feature emerged from L2 (see inset in Fig.~\ref{Lightcurve}(e) and Fig.~\ref{DEM}(d,e)), reaching its maximum altitude at $[x,y]=[1045'',-45'']$. The EM distribution of a position along the jet-like feature before, during and after its occurrence is shown in Fig.~\ref{DEM}(f) (see position and corresponding times marked in Fig.~\ref{DEM}(e)), showing that it was composed of mostly hot plasma above 2~MK, with a major contribution coming from plasma at $\sim$6~MK (compare EM of light and dark blue lines at Log T = 6.85). Co-temporal with the jet-like feature reaching its maximum altitude was the occurrence of localized heating (see white arrow in Fig.~\ref{DEM}(c)). 

\subsection{Characteristics of the cusp-shaped flare loops}
     \label{S-Characteristics of the cusp-shaped flare loops}

The impulsive and decay phases revealed the unusual static cusp-shaped appearance of the flare loops, seen in \aia\ 94 and 131 \AA. The loops reach an approximate altitude of 66~Mm above the solar surface, connecting the precursor locations (see Fig.~\ref{Lightcurve}(e,f)).

A fast decrease of the SXR emission occurred during the first 6~minutes of the decay phase, \ie\ $\sim$21:51--21:56~UT (see red and black lines in Fig.~\ref{Lightcurve}). The \rhessi\ spectral analysis revealed non-thermal X-ray emission from the high cusp-shaped flare loops (Fig.~\ref{spectra}(f)). X-ray analysis during the rest of the decay phase was not possible since \rhessi\ entered the Southern Atlantic Anomaly (SAA) and then night time. However, the \goes\ SXR emission did not drop back down to background level. On the contrary, an extended period of enhanced SXR flux followed (above $1.5\cdot10^{-6}~Wm^{-2}$), indicative of additional energy release during the decay phase (see black curve in Fig.~\ref{Lightcurve}). Similar behavior is exhibited by the \aia~94~\AA\ lightcurve, indicative of further energy input and heating (see green curve in Fig.~\ref{Lightcurve}). Additionally, enhanced SXR emission in the 6--12~keV energy band in between the SSA and \rhessi's night time is registered. We note that, a \rhessi\ image during this period was not possible due to the low number of counts.

\begin{figure*}[ht]
\centerline{
\centering\includegraphics[width=0.85\textwidth]{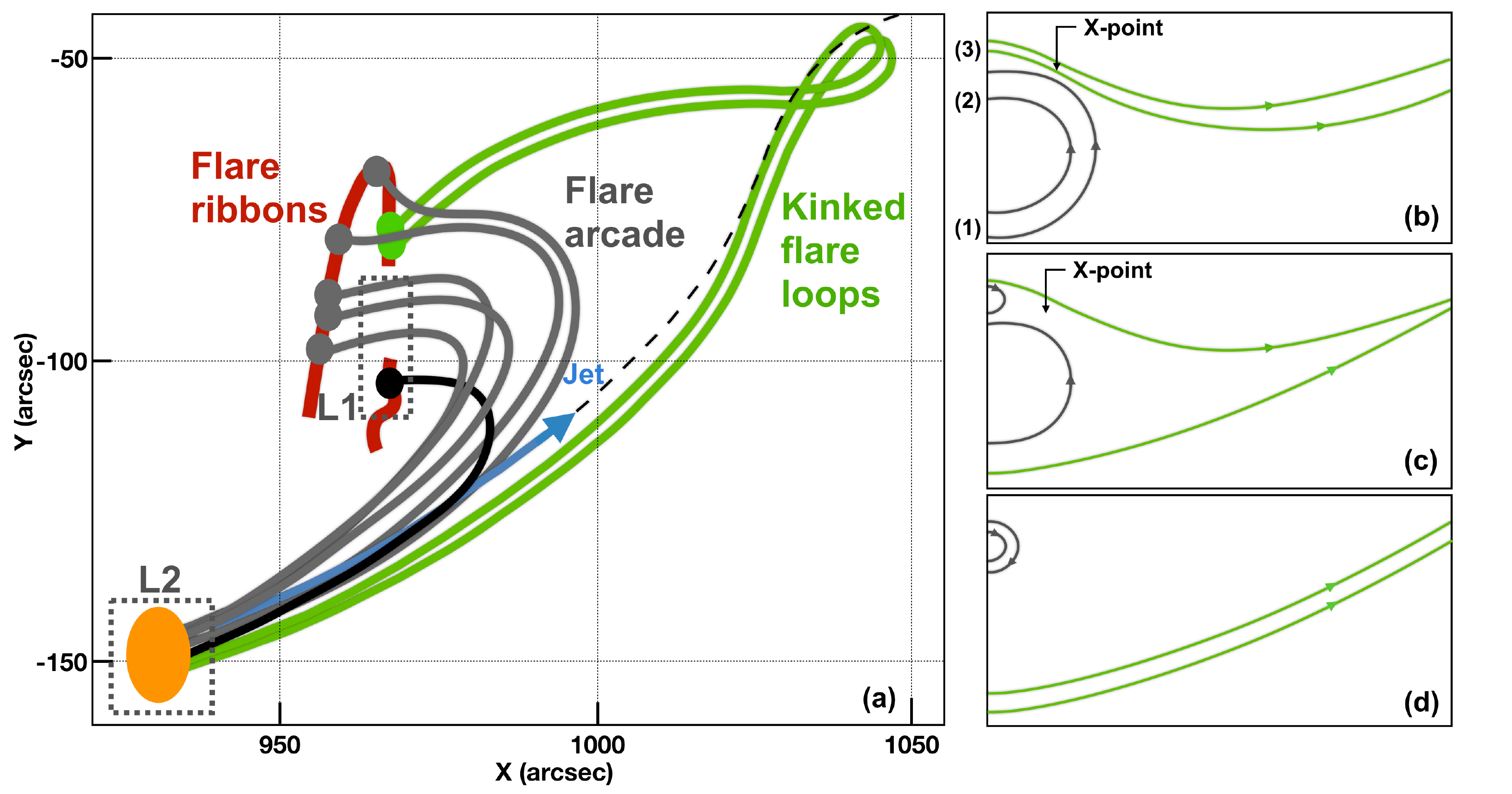}
}
\caption{Diagram of the magnetic configuration of the flare. The grey lines represent the flare arcade, the black line is the loop connecting L1 and L2 (see Fig.~\ref{EUV_summary}(d)). The green lines represent the kinked flare loops. The blue arrow represents the jet--like feature (see Fig.~\ref{Lightcurve}(e)). Insets [1--3] represent the most probable scenario of interchange reconnection in which the flare was triggered at L2 and the connectivity with the higher corona was established.\\
}
\label{sketch}
\end{figure*}

(E)UV imaging reveals ongoing heating of the cusp-shaped flare loops while the underlying flare arcade cools down (Fig.~\ref{Lightcurve}(g,h)). This led to an unexpected temperature distribution (see Fig.~\ref{DEM}(d,f) and animation), \ie\ the top of the cusp-shaped flare loops was significantly hotter ($\sim$10--12~MK) than the arcade underneath ($\sim$4--6~MK). The temperature of the cusp from the DEM analysis is consistent with the temperature of the thermal plasma derived from the \rhessi\ spectra, of about 10~MK (see spectra associated with Fig.~\ref{spectra}(f)). 

\section{Discussion and Conclusion} 
      \label{S-Discussion and Conclusion}

The study of SOL2014-01-13T21:51M1.3 poses challenges to previously reported flare models. Not only was it confined, it exhibited cusp-shaped flare loops, a typical feature of eruptive flares. We studied the early, impulsive and decay phases in order to understand the initiation and development of such an uncommon structure in confined flares.

(E)UV observations show a direct relationship between the last precursors (P2, P3) and the triggering of the flare, as they were co-spatial and occurred subsequently (Fig.~\ref{Lightcurve}). Hard X-ray analysis of the precursors revealed the non-thermal nature of P2 and P3 (Fig.~\ref{spectra}(b,c)), indicative of accelerated electrons prior to the flare onset \citep[in agreement with \eg][]{2018ApJ...854...26L,2019ApJ...874..122H}. This suggests that reconfiguration of the magnetic field due to pre-flare reconnection caused the strong magnetic reconnection and triggered SOL2014-01-13T21:51M1.3 at L2. This is supported by (a) the appearance of an additional bright loop during the early impulsive phase (see inset of Fig.~\ref{Lightcurve}(d)) that takes part in the reconnection process that, which probably established new connectivity between the lower and higher corona; (b) the strong non-thermal source imaged during the impulsive phase (Fig.~\ref{spectra}(d,e)) revealing that L2 is the compact main flare region; and (c) the jet-like feature originating at L2 during the impulsive phase (see inset in Fig.~\ref{Lightcurve}(e)).

The triggering of SOL2014-01-13T21:51M1.3 finally exposed its most intriguing observational features: the cusp-shaped appearance of the high-lying flare loops, their prolonged hot emission and the increasing temperature distribution with height, \ie\ all signatures typical of eruptive events \citep{1992PASJ...44L..63T}. Given the observational characteristics presented above, the most plausible magnetic scenario is depicted in Fig.~\ref{sketch}.

We observed a jet-like feature emerging from L2 during the impulsive phase. Jets usually occur due to interchange reconnection \citep{2002JGRA..107.1028C}, \eg\ reconnection between closed and open fields that enable the emergence of chromospheric material along the newly reconnected fields \citep{1989ApJ...345..584S,1992PASJ...44..265S,1996PASJ...48..353Y}. The jet-like feature reported here was most probably generated as a consequence of the reconnection process that established the connectivity between L2 and the higher corona (\ie\ reconnection between closed loops of different scales) in a manner similar to interchange reconnection. A simplified sequence depicting this process is represented in Fig.~\ref{sketch}(b--d), showing small loops from L2 (in grey) reconnecting with much larger loops (in green) that establish the connectivity with the higher corona. On the other hand, although the jet-like feature is composed mostly of hot flaring plasma, the low temperature component at transition region temperatures around 0.3-0.5 MK indicates that it could consist of plasma from both the reconnection outflow and chromospheric evaporation from the flare footpoints along the newly reconnected large scale fields (due to the impact of high energy particles once the connectivity between the lower and higher corona was established). The complex dynamics of chromospheric evaporation flows during flares, microflares and jets, in particular in transition region lines, has, \eg, been reported in \cite{2010ApJ...719..655V} and \cite{2009A&A...505..811B}.

The heating of the higher corona was initiated by the injection of hot plasma from L2, as suggested by the quick response of the cusp-shaped flare loops during the impulsive phase (Fig.~\ref{DEM}(e)). The nature of the static high-lying hot flare loops of SOL2014-01-13T21:51M1.3 was clearly different from what has been previously reported in literature \citep{2014ApJ...790....8L,2015SoPh..290.2211G}. The cusp-shaped appearance of the flare loops of SOL2014-01-13T21:51M1.3 (see Fig.~\ref{Lightcurve}(e--h)) is most probably due to a kink of the flare loops (see green lines in Fig.~\ref{sketch}). The \rhessi\ sources (of a partly non-thermal nature) near the projected crossing point of the kinked structure during the decay phase (see Fig.~\ref{spectra}(f)), provide evidence of energy release occurring as a consequence of magnetic reconnection at the kink. The extended enhanced SXR and EUV emission during the late phase of SOL2014-01-13T21:51M1.3 provides more evidence of a weak process of ongoing energy release (see Fig.~\ref{Lightcurve}). 

X-ray emission at the crossing point of a kinked flux rope as a consequence of magnetic reconnection has been previously imaged in eruptive \citep[\eg][]{2009ApJ...697..999L,2009ApJ...703....1C,2012ApJ...746...17G} and confined events \citep[\eg][]{2006ApJ...653..719A}. Whether or not the reconnection event is enough to produce an ejection depends on the degree of twist of the flux rope \citep[\eg][]{2010SoPh..266...91K}. However, the nature of the kinked fields in the flare under study differs from the works in the previous studies in that they are not part of a low-lying flux rope but rather enveloping the flare arcade.

We therefore interpret the observed prolonged hot emission and temperature distribution as a consequence of slow magnetic reconnection caused by a strong increase of the thermal pressure at the kink. This can explain the continuous energy release during the late phase revealed by the prolonged SXR emission (see Fig.~\ref{Lightcurve}), and the increasing temperature distribution with height (see Fig.~\ref{DEM}(d)). This magnetic configuration also explains why, in this case, we do not observe a growing arcade, a signature of reconnection progressively occurring higher up in the corona, as predicted by the standard flare model.

The answer to an important question remains elusive, \ie\ was the kink formed as a consequence of the reconnection at L2 or did it pre-exist before the flare onset? In the first scenario, the flare dynamics initiated by the interchange reconnection at L2 were responsible for the formation of both, the overlying kinked loops as well as the low-lying flare arcade. A possible explanation for the kinked overlying field could involve the transfer of twist in the course of the jet \citep[\eg][]{1986SoPh..103..299S}. In the second scenario, the overlying field was kinked prior to the flare onset and heated by accelerated particles from L2 and slow magnetic reconnection at the helical current sheet, formed at the interface with the surrounding plasma \citep[\eg][]{2004A&A...413L..23k}. The low-lying flare arcade may then have been produced by reconnection at a current sheet, presumably underlying the kinked structure.

Our study underlines the complexity of flare processes in that features typical of both confined and eruptive flares can be observed in a single event. This must be accounted for by any realistic flare model.

\acknowledgments
\small

We thank the anonymous referee for the helpful comments that significantly helped to improve the manuscript. A.H. A.M.V. and J.K.T acknowledge the Austrian Science Fund (FWF): P27292-N20. Y.S. acknowledges the Major International Joint Research Project (11820101002) of NSFC, the Joint Research Fund in Astronomy (U1631242, U1731241) under the cooperative agreement between NSFC and CAS, the CAS Strategic Pioneer Program on Space Science, (grant No. XDA15052200, XDA15320300, XDA15320301), the Thousand Young Talents Plan, and the Jiangsu ``Double Innovation Plan". This work is supported by the Indo--Austrian joint research project no. INT/AUSTRIA/BMWF/ P-05/2017 and OeAD project no. IN 03/2017. K.D. acknowledges funding by the Austrian Space Applications Programme of the Austrian Research Promotion Agency FFG (ASAP-14 865972, SSCME). A.H. and J.K.T acknowledge Dr. D. Utz for helpful discussions of this event. \sdo\ data are courtesy of the NASA/\sdo\ and HMI science team. \rhessi\ is a NASA Small Explorer Mission. \goes\ is a joint effort of NASA and the National Oceanic and Atmospheric Administration (NOAA).



\bibliography{bibliography}   

\end{document}